\documentclass[aps,prb,reprint,showpacs,superscriptaddress]{revtex4-1}
\usepackage{graphicx}
\usepackage{amsmath}
\usepackage{amssymb}
\usepackage{amsfonts}
\usepackage{dcolumn}
\usepackage{dsfont}
\usepackage{latexsym}
\usepackage{rotating}
\usepackage{color}
\usepackage{latexsym}
\usepackage{bbm}
\usepackage{subfigure}
\usepackage{float}
\usepackage{epsfig}
\usepackage{psfrag}
\usepackage{bm}
\usepackage{amsthm}
\usepackage{eucal}
\usepackage{url}
\usepackage{natbib}
\usepackage{mathtools}
\DeclarePairedDelimiter{\abs}{\lvert}{\rvert}

\usepackage{color} 

\DeclareMathAlphabet\mathbfcal{OMS}{cmsy}{b}{n}

\usepackage{hyperref}
\hypersetup{
colorlinks=true,final=true,
        linkcolor=blue,
        citecolor=blue,
        filecolor=blue,
        urlcolor=blue,
}

\begin{document}

\title{Body centered phase of Cu at high temperature and pressure}
\author{Urmimala Dey}
\affiliation{Centre for Theoretical Studies, Indian Institute of Technology, Kharagpur-721302, India}
\author{Nilanjan Mitra}
\affiliation{Centre for Theoretical Studies, Indian Institute of Technology, Kharagpur-721302, India}
\affiliation{Civil Engineering Department, Indian Institute of Technology, Kharagpur-721302, India}

\author{A. Taraphder}
\affiliation{Centre for Theoretical Studies, Indian Institute of Technology, Kharagpur-721302, India}
\affiliation{Department of Physics, Indian Institute of Technology, Kharagpur-721302, India}
\affiliation{School of Basic Sciences, Indian Institute of Technology, Mandi, HP 175005 India}



\begin{abstract}
{The existence of a body centered tetragonal phase of Cu has been investigated in this manuscript when the single crystal FCC Cu is subjected to both high pressure and high temperature. The results have been demonstrated through DFT calculations (which are typically done at 0 K) followed by Helmholtz free energy calculations (for high temperature). The new metastable phase of Cu demonstrates higher thermal conductivity compared to that of the FCC phase and thereby may be beneficial for high temperature engineering applications. 
}
\end{abstract}

\maketitle

\section{Introduction}
Landau's theory of phase transition states that different phases of a material may be stable when it is subjected to different types of external imposed constraints which could be temperature, pressure, magnetic field, concentration degree of cross linking or any number of other physical quantities. Typically, the Landau free energy functional is assumed to be function of external constraint and an order parameter which undergoes a discontinuity whenever there is a first order phase transition. For the case of crystal structures, the primary parameter upon which the Landau energy functional depends is on the mass density $\rho$(\textbf{r})
\begin{equation}
\rho(\textbf{r}) = \bar{\rho} + \sum_{\textbf{q} \in \textbf{G}} \bigg[ \rho(\textbf{q})e^{-i \textbf{q} \cdot \textbf{r}} + c.c. \bigg]
\end{equation}
where $\bar{\rho}$ is the mean density and \textbf{G} is the set of reciprocal vectors that characterize the crystal structure. The wavenumber is given as $q = 2\pi/\lambda$ where $\lambda$ represents the wavelength. A complex conjugate (c.c) is added to retain a real number for mass density. It can be adjudged from the above formulae that solid-solid phase transition in a crystal structure is dependent upon phonon vibrations.

Typically Cu exists as a face-centered-cubic material at ambient temperature and pressure conditions. The non-existence of body centered phase of Cu at ambient pressure and 0K temperature has been proved by numerous researchers through DFT based studies.~\cite{Lu_1990,Kraft_1993} However, it is interesting to observe that in molecular beam epitaxy, BCC films of Cu have been grown pseudomorphically on Pd\{001\}, Pt\{001\}, Ag\{001\} and Fe\{001\} substrate.~\cite{Wang_1987,Li_1989,Li_1991_1,Li_1991_2} Typically in beam epitaxy, the sample is subjected to a constrained pressure (which in this case is induced by the substrate grain boundary). The controversy of unstable existence of body centered phase of Cu at ambient pressure and temperature conditions along with experimental observations of body centered phase of Cu in epitaxy was eventually resolved. The explanation provided was that the stable substrates, acting as grain boundaries, influence the phase change in Cu and it can happen only in thin films and not for bulk material or even thick films.~\cite{Wang_1999} An existence of a BCT
phase of Cu with c/a = 0.93 was demonstrated~\cite{JonaMarcus} which is observed to be tetragonally stable by  calculation  of  the  epitaxial  Bain  path  of  Cu.   However, it has also been mentioned in the manuscript since this  special  BCT  phase  does  not  satisfy  all  the  stability  criteria  imposed  on  elastic  constants,  it  is  unstable against other modes of shear deformation. Typically,
these ab-initio studies have been carried out at ground state; there maybe a possibility of stable body-centered phase of Cu at higher temperatures and pressures which is yet to be explored.

Apart from molecular beam epitaxy studies which typically constrains the lattice structure by application of pressure from substrate grain boundaries, Cu-based shape memory alloys also exhibit a BCC phase at high temperatures. It has been reported through Neutron diffraction studies that body centered phase of Cu is present in Cu based shape memory alloys such as in Cu-Zn-Al~\cite{Guenin_1979}, Cu-Al-Ni~\cite{Hoshino_1975}, Cu-Al-Pd,~\cite{Nagasawa_1992}, Cu-Al-Be~\cite{Manosa_1993} in which the whole TA2[110] phonon branch was observed to soften with temperature as transition temperature is approached. It should be mentioned in this regard that it is well known that entropy changes in a Martensitic Transformation is due to contributions from vibrational component of the crystal lattice and electronic contributions near the Fermi surface. Friedel~\cite{Friedel} postulated that even though body centered phase of Cu is energetically unstable at the ground state, it may be the preferred system at high temperature due to its large entropy resulting from low-energy vibrational transverse modes. For Cu based  shape  memory alloys it was observed through calorimetric and magnetic measurements that harmonic vibration of the lattice (specifically the coupling between homogenous shear and short wavelength phonon) is the main reason for stability of the BCC phase at high temperature and the electronic contribution to entropy change is negligible.~\cite{Planes_1992,Planes_1993,Planes_1996} Comparing this low lying phonon branch to Zener elastic modes,~\cite{Zener_1947} a new Hamiltonian has also been proposed~\cite{MorrisGooding_1990} which displays vibrational-entropy-driven first-order solid-solid diffusionless martensitic phase transitions.  The model Hamiltonian is suitable for high temperature applications since it employs anharmonic intersite couplings which alters the vibration stiffness with changes in temperature.  

It should be realized that in Cu-based soft memory alloys, the Cu crystal structure is constrained (due to presence of other materials eventually resulting in development of pressure) which undergoes a phase transition to that of a body centered phase on high temperature. Thereby, given this observation it is quite conceivable that under high temperature and pressure conditions there may exist a body centered phase of Cu. Body centered tetragonal phase of Cu was also demonstrated to develop when Cu single crystals are subjected to shock compression along the [100] direction at a piston velocity of around 1.5 $-$ 2 km/s.~\cite{Neogi_2017} 

This present study is aimed at investigating the possibility of existence of a metastable body centered tetragonal phase of Cu at high temperature and pressure. The study also demonstrates changes in properties of the material that are associated with phase transformation of Cu from FCC to BCT phase; which might have future engineering applications.  

\section{Simulation methodology}
We perform density functional theory calculations using the full potential linearized augmented plane wave (FLAPW) method implemented in the WIEN2K package.~\cite{Blaha_2002} The generalized gradient approximation (GGA) of Perdew-Burke-Ernzerhof (PBE)~\cite{PBE_1996} is employed for the exchange correlation part. A k-mesh of 20 $\times$ 20 $\times$ 20 k-points is used for the whole Brillouin zone. We use $L_{max}$ = 12 for the expansion of partial waves and $G_{max}$ = 14 for the charge Fourier expansion. The muffin tin radii (RMT) and $K_{max}$ are chosen such $RMT \times K_{max}$ = 7.0, where $K_{max}$ is the largest plane wave vector used in the plane-wave expansion.

Phonon dispersion spectra and Helmholtz free energies are calculated within the framework of quasi-harmonic approximation using the finite displacement method in the PHONOPY code.~\cite{Togo_2008} For calculating the real space force constants we use the projector augmented wave (PAW) method as implemented in the Vienna Ab initio Simulation Package (VASP).~\cite{Kresse_1996} We take an energy cutoff of 500 eV for the plane waves and a 11 $\times$ 11 $\times$ 11 Monkhorst-Pack k-mesh~\cite{Monkhorst_1976} is adopted to integrate the full Brillouin zone. The atomic positions are relaxed until the maximum Hellmann-Feynman forces on the atom are smaller than 1 meV/\r{A}. Energy convergence criterion is set to $10^{-8}$ eV. 

We determine the elastic constants by analyzing the changes in the calculated stress values resulting from changes in the strain as implemented in the VASP code. In the stress-strain approach, a set of linear equations are constructed from the stress tensors calculated from each deformation and the solutions of the linear system of equations are found by using orthogonal matrix factorizations. This approach uses the well-known tensorial form of the Hooks law, that describes the following relation between the stress component and the applied strain ~\cite{Mouhat_2014}. The symmetric elastic constants and are calculated after fully relaxing a crystal structure. 

Transport properties are determined with the BoltzTraP code~\cite{Madsen_2006} interfaced with WIEN2K using a dense k-mesh of 30 $\times$ 30 $\times$ 30 k-points. In order to obtain an analytical expression of bands, BoltzTraP code depends on a well tested smoothed Fourier interpolation. The constant relaxation time approximation is used in the calculations. Since the electrons contributing to transport reside in a narrow energy range due to the delta-function like Fermi broadening, the relaxation time can be assumed to be nearly the same for all the electrons. In our DFT calculations, the temperature dependence of the electronic band structure is ignored.
\section{Results}
\begin{figure}[h]
\centering
$\begin{array}{ccc}
\includegraphics[scale=.34]{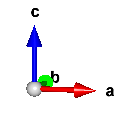}  &
\includegraphics[scale=.25]{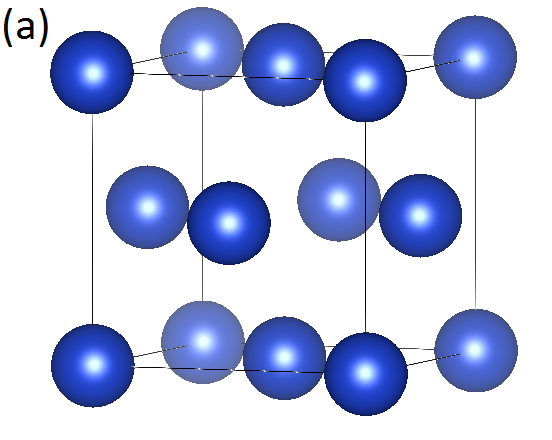}  &
\includegraphics[scale=.24]{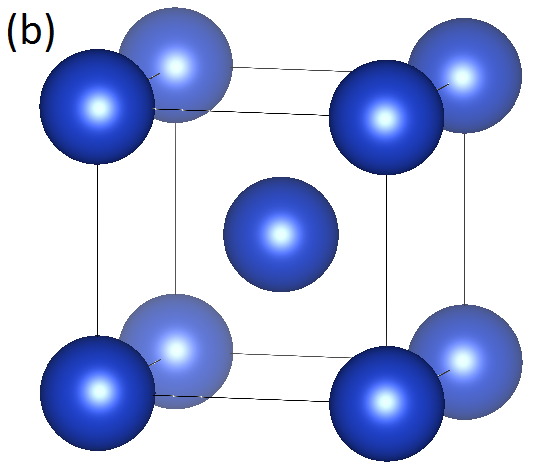} \\
\end{array}$
\caption{Crystal structures of the (a) FCC (space group Fm$\bar{3}$m) and the (b) BCT (space group I4/mmm) phases of Cu.}
\label{structure}
\end{figure}

It is well known that at ambient pressure and temperature, Cu crystallizes in a face-centred cubic (FCC) structure with space group symmetry Fm$\bar{3}$m (no. 225), as shown in Fig.~\ref{structure}(a). 

In order to obtain the equilibrium lattice parameters, we optimize the total energy of the system as a function of volume and fitted the data to the Birch-Murnaghan (B-M) equation of state,~\cite{Murnaghan_1937,Birch_1947} 
\begin{figure}[h]
\centering
\includegraphics[scale=.23]{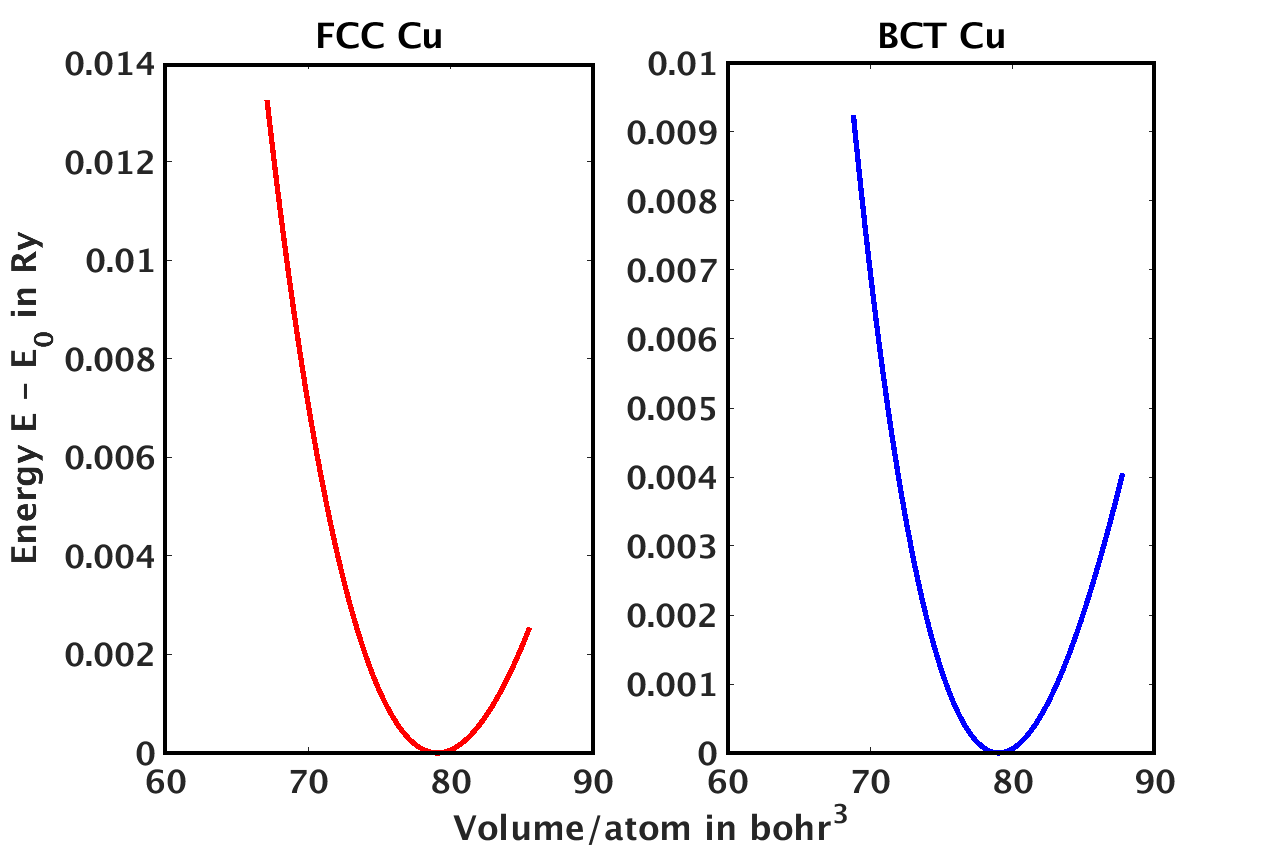}  
\caption{Total energy of the FCC (left) and BCT (right) phases as a function of volume at ambient pressure.}
\label{E_V}
\end{figure}
\begin{equation}
\begin{split}
P(V) &= \frac{3B}{2} \Bigg[{\Bigg(\frac{V_0}{V}\Bigg)}^\frac{7}{3} - {\Bigg(\frac{V_0}{V}\Bigg)}^\frac{5}{3} \Bigg] \\
&\times \Bigg\{ 1 + \frac{3}{4}(B^\prime- 4)\Bigg[{\Bigg(\frac{V_0}{V}\Bigg)}^\frac{2}{3} - 1\Bigg] \Bigg\}
\end{split}
\end{equation}
where, $P$, $V_0$, $V$, $B$, $B^\prime$ denote the pressure, reference volume, deformed volume, bulk modulus and the pressure derivative of the bulk modulus respectively.
\begin{figure}[h]
\centering
\includegraphics[scale=.23]{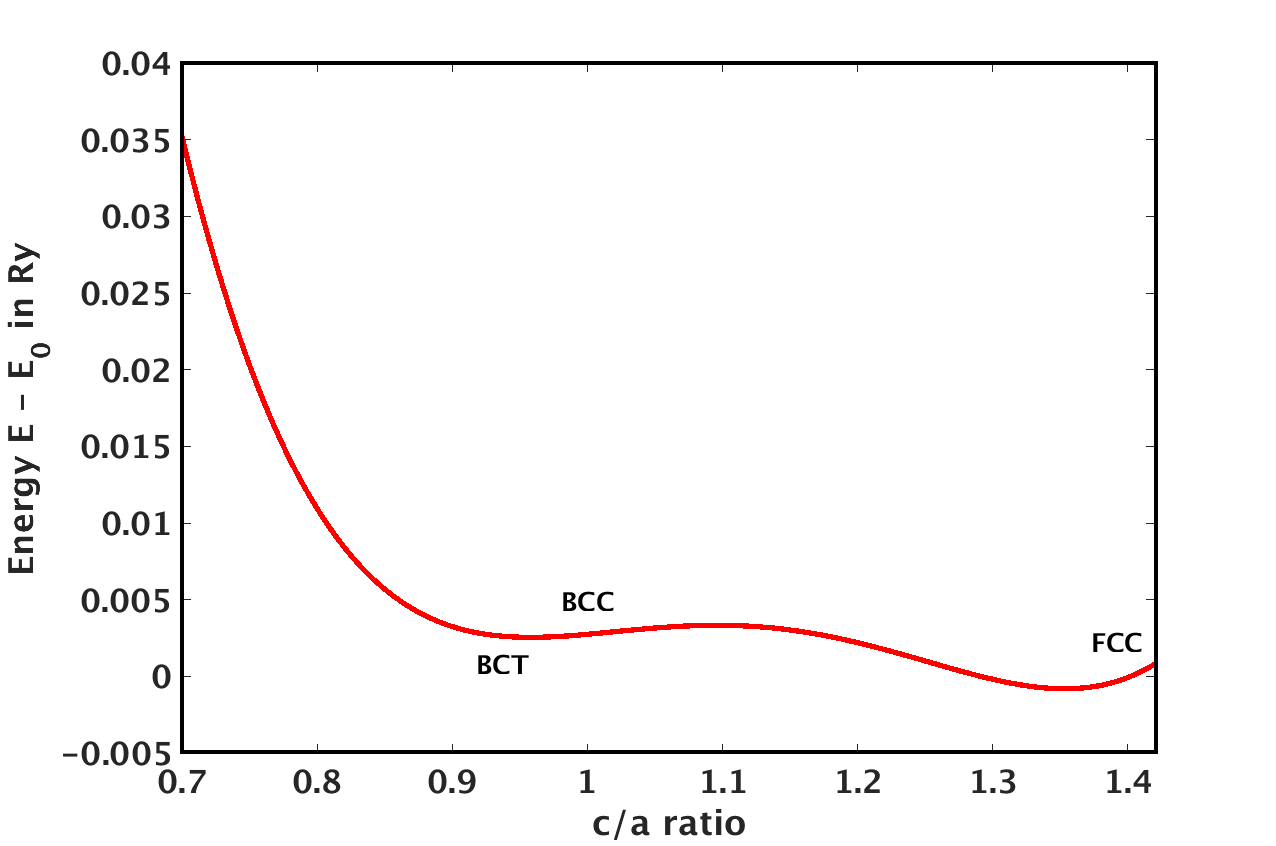}  
\caption{DFT calculated energy of the different phases of Cu versus $\frac{c}{a}$ ratio at high pressure ($\sim$ 80 GPa). At high pressure, a local minimum appears for $\frac{c}{a}$ $\sim$ 0.966. Energy of this high pressure BCT phase is lower than the BCC phase. However, the global minimum corresponds to the FCC phase.}
\label{c_a}
\end{figure}
\begin{figure*}
\onecolumngrid
\centering
$\begin{array}{cc}
\includegraphics[scale=.419]{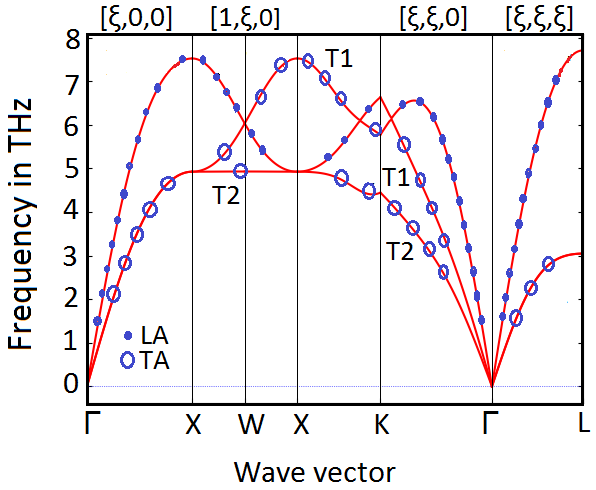}  &
\includegraphics[scale=.42]{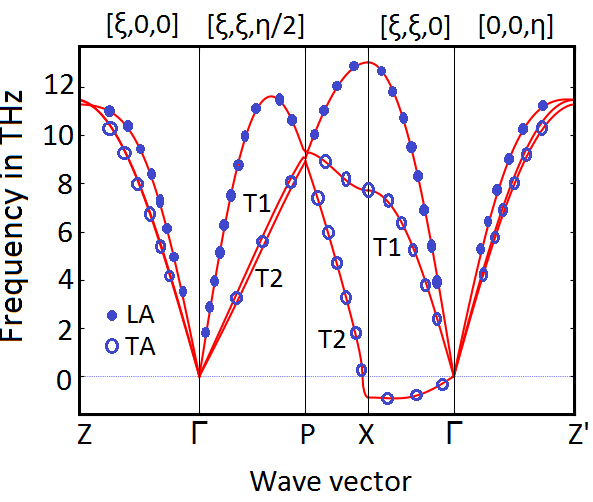} \\
\end{array}$
\caption{Phonon band dispersions for the known FCC (left) phase and the high pressure BCT (right) phase of Cu. The filled circles denote the LA modes and the open circles show the TA modes. LA and TA are the longitudinal and transverse acoustic modes respectively. T1 and T2 are the two TA modes. In the BCT phase, softening of phonon modes along the $[\xi \xi 0]$ direction indicates the dynamical instability of the BCT structure at low temperature (0K) and high pressure.}
\label{phonon}
\twocolumngrid
\end{figure*}

At ambient pressure, we find that the minimum energy structure of the FCC phase corresponds to the equilibrium lattice parameter of $a$ = 3.605 \r{A}. Again at ambient pressure, the total energy optimization with respect to the $\frac{c}{a}$ ratio shows that for the BCT phase (space group I4/mmm, no. 139), the minimum occurs at the FCC structure with $\frac{c}{a}$ $\sim$ 1.414. Keeping $\frac{c}{a}$ fixed at 1.414, we minimize the total energy of the BCT structure as a function of volume (Fig.~\ref{E_V}) from which we obtain $a$ = 2.550 \r{A} and $c$ = 3.605 \r{A}, which match well with the previous first-principles total energy calculations of BCT Cu by Morrison \textit{et al}.~\cite{Morrison_1989} The crystal structure of the BCT phase is shown in Fig.~\ref{structure}(b). 
   
However, when it is subjected to high pressure, the FCC structure becomes unstable with respect to a tetragonal distortion and at shock pressure of $\sim$ 80 GPa, the FCC structure is transformed into a body centred tetragonal (BCT) structure.~\cite{Neogi_2017} This FCC-BCT phase transition results in the appearance of a local minimum in the total energy versus $\frac{c}{a}$ ratio curve, as shown in Fig.~\ref{c_a}. In Fig.~\ref{c_a}, we plot the total energy as a function of the $\frac{c}{a}$ ratio, keeping the volume of the cell fixed at 8.993 ${\text{\r{A}}}^3$/atom corresponding to $\sim$ 80 GPa pressure. We find that the local minimum in the total energy corresponds to $\frac{c}{a}$ $\sim$ 0.966, which is consistent with the previously reported value of $\frac{c}{a}$ obtained in the shock wave study of single crystal Cu.~\cite{Neogi_2017} The calculated equilibrium lattice parameters are $a$ = 2.650 \r{A} and $c$ = 2.560 \r{A}. 

In order to check the mechanical stability of the high pressure BCT phase found, we have calculated the phonon spectrum of the BCT structure using the equilibrium lattice parameters and compared it with the phonon spectrum of the FCC phase. As seen from Fig.~\ref{phonon}, the transverse acoustic (TA) modes are hardened in the BCT phase along the $Z$-$\Gamma$ and $\Gamma$-$Z'$ directions. On the contrary, a soft mode appears along the $[\xi \xi 0]$ direction. In case of Cu-based shape memory alloys, it is found that the low energy of the TA2 [110] phonon mode contributes significantly to the excess of entropy which stabilizes the tetragonal phase at high temperature~\cite{Planes_1992} and has been proved experimentally.~\cite{Planes_1992} However, the softening of the TA2 mode near the $X$-point indicates that the shear modulus, obtained by taking the derivative of the TA2 mode will be negative and will lead to the mechanical instability of the BCT phase at high pressure.

The calculated elastic constants of the FCC and BCT structures are shown in Table~\ref{tab1}. For cubic systems, the elastic matrix $c_{ij}$ has only three independent components :  $c_{11}$, $c_{12}$ and $c_{44}$ and the Born stability criteria for the mechanical stability of cubic systems are given by~\cite{Mouhat_2014} 
\begin{subequations}
\label{FCC_SC}
\begin{equation}
c_{11} - c_{12} > 0 
\end{equation} 
\begin{equation}
c_{11} + 2c_{12} > 0
\end{equation}
\begin{equation}
c_{44} > 0 
\end{equation}
\end{subequations}
\begin{table}[h]
\centering
\caption{Elastic constants of the FCC and BCT phases of Cu at T = 0 K. Since the shear modulus $c^\prime$ is negative for the BCT phase, the BCT structure is unstable with respect to shear deformations and will be mechanically unstable.}
\label{tab1}
\begin{tabular}{|c|c|c|}
\hline
    Elastic constants in GPa & FCC Cu & BCT Cu \\
    \hline
    $c_{11}$  & 230.81 & 400.60 \\
    $c_{12}$  & 119.00 & 445.16\\
    $c_{13}$  & $-$      & 376.98 \\
    $c_{33}$  & $-$      & 488.77 \\
    $c_{44}$  & 114.01 & 267.47 \\
    $c_{66}$  & $-$ & 278.43  \\
    $c^\prime$ & 55.91 & $-$22.28 \\
    \hline
  \end{tabular}
\end{table}

However, in case of tetragonal systems, there are six independent elastic constants : $c_{11}$, $c_{12}$, $c_{13}$, $c_{33}$, $c_{44}$ and $c_{66}$, which satisfy the following necessary and sufficient conditions for the stability of a tetragonal system~\cite{Mouhat_2014} :
\begin{subequations}
\begin{equation}
c_{11} > \abs{c_{12}}  \label{BCT_SC1}
\end{equation} 
\begin{equation}
2c_{13}^2 < c_{33}(c_{11} + c_{12}) \label{BCT_SC2}
\end{equation}
\begin{equation}
c_{44} > 0 \label{BCT_SC3}
\end{equation}
\begin{equation}
c_{66} > 0  \label{BCT_SC4}
\end{equation}
\end{subequations}
Our DFT calculated elastic constants for the FCC phase, as listed in Table~\ref{tab1}, fulfill all the stability criteria (Eq.~(\ref{FCC_SC})), which implies that the FCC phase is stable at ambient pressure and low temperature. By contrast, the elastic constants of the high pressure BCT phase satisfy the conditions~(\ref{BCT_SC2}),~(\ref{BCT_SC3}) and (\ref{BCT_SC4}), but fail to satisfy condition~(\ref{BCT_SC1}). Therefore, the shear modulus $c^\prime$ = $\frac{c_{11} - c_{12}}{2}$ $<$ $0$ (Table~\ref{tab1}), suggesting that the BCT structure would decrease its energy by deformations of the cell and as a result the BCT phase will be unstable at low temperature (0K).
\begin{figure}[h]
\centering
\includegraphics[scale=.24]{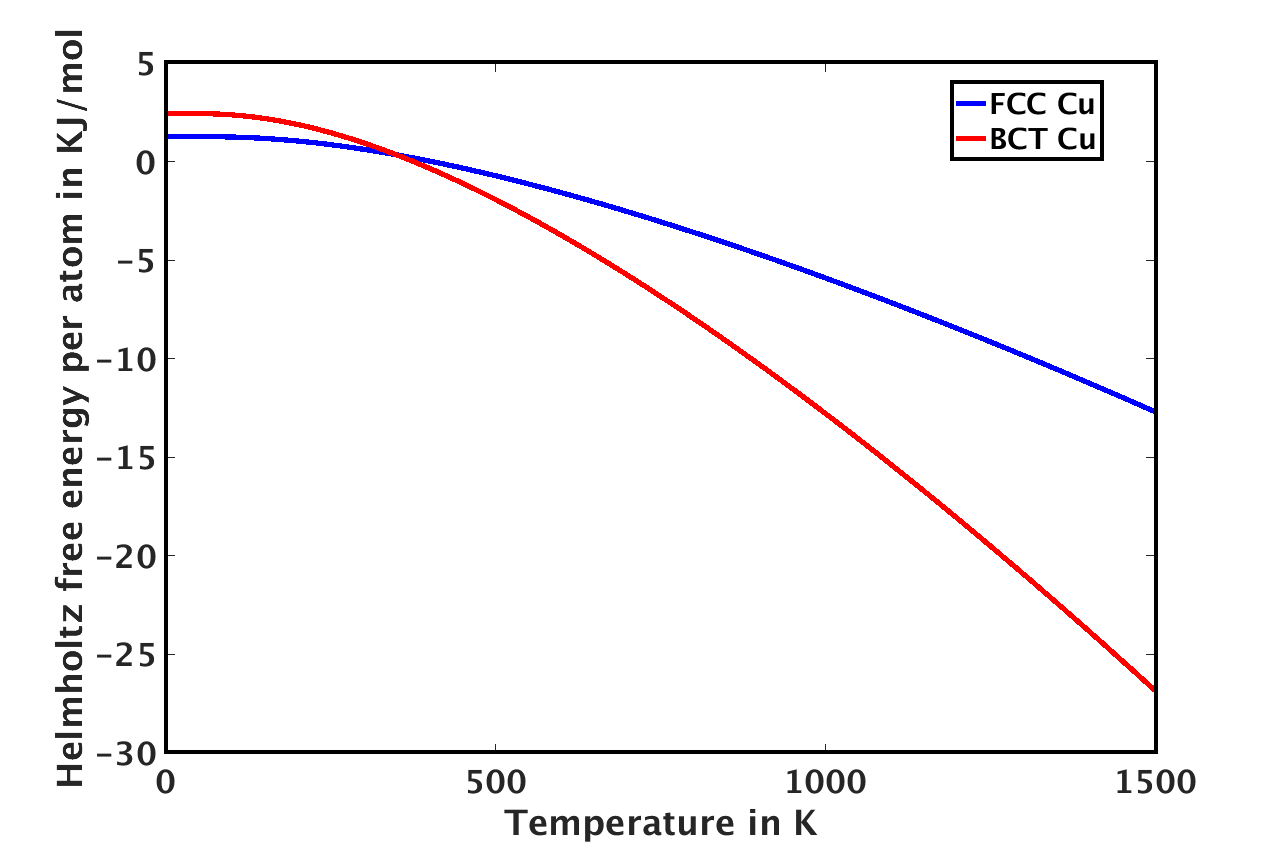}  
\caption{The Helmholtz free energy per atom of the two phases of Cu as a function of temperature at $\sim$ 80 GPa. At low temperature, the FCC phase is more stable than the BCT phase. However, as the temperature rises, the BCT phase gains stability over the FCC phase.}
\label{free_energy}
\end{figure}
\begin{figure}[t]
\centering
$\begin{array}{cc}
\includegraphics[scale=.20]{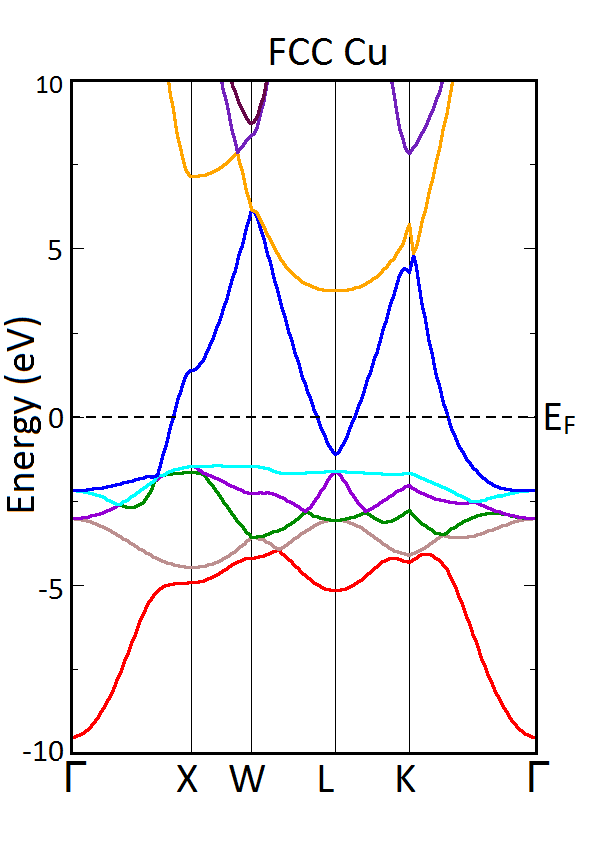}  &
\includegraphics[scale=.20]{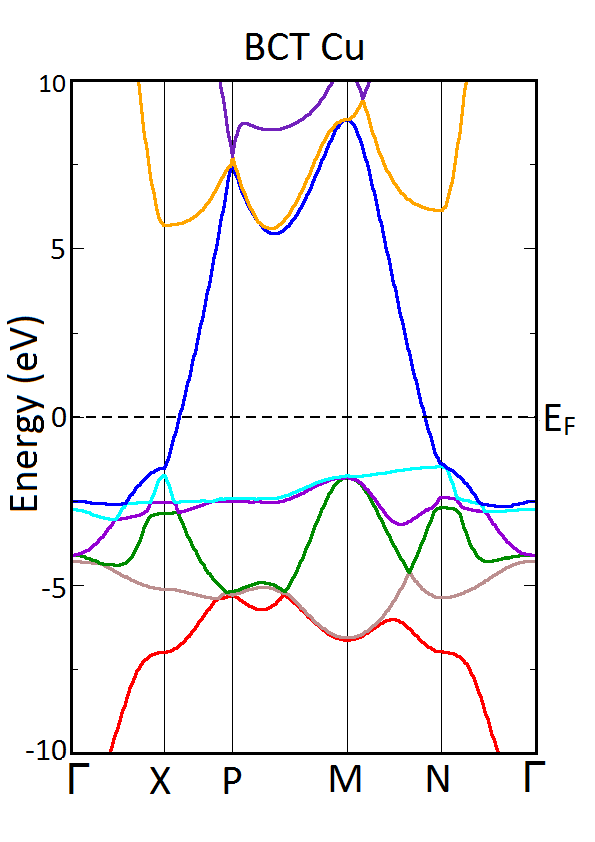} \\
\end{array}$
\caption{Electronic band structures of the FCC (left) and BCT (right) phases of Cu. Since there are finite density of states near the Fermi level, both the structures are metallic.}
\label{band_structure}
\end{figure}

However, the dynamical instability of a crystal structure at low temperature does not mean that the crystal structure is unstable at high temperature. Since our \textit{ab initio} calculations are carried out at ground state ($T$ = 0 K), there may be a possibility that the BCT phase may attain stability at high temperature. For example, the body centred cubic (BCC) phase of Fe gains stability at high temperature,~\cite{Anderson_1997,Ahuja_2003,Vocadlo_2003} though it is not stable at high pressure and low temperature ($T$ = 0 K). From Fig.~\ref{c_a} we find that at high pressure, the BCT phase results in the appearance of a local minimum in the energy vs. $\frac{c}{a}$ ratio plot, however, the energy of the BCT phase is higher than the FCC minimum, indicating that the FCC phase is more stable compared to the BCT phase at low temperature and high pressure ($\sim$ 80 GPa). In order to check the relative stability of the two phases (FCC and BCT) at higher temperatures, we calculate the Helmholtz free energy (per atom) of the two phases employing the information of the phonon density of states at the volume 8.993 ${\text{\r{A}}}^3$/atom. From Fig.~\ref{free_energy}, it is found that as the temperature rises from 0 K, the FCC phase becomes unstable with respect to the BCT structure and at $\sim$ 1130 K, the temperature of the shock wave,~\cite{Neogi_2017} the difference in the free energy of the BCT and the FCC phase is $\sim$ -8.64 KJ/mol/atom. This suggests that the BCT phase gains stability over the FCC phase with increasing temperature. 
\begin{figure}[h]
\centering
\includegraphics[scale=.24]{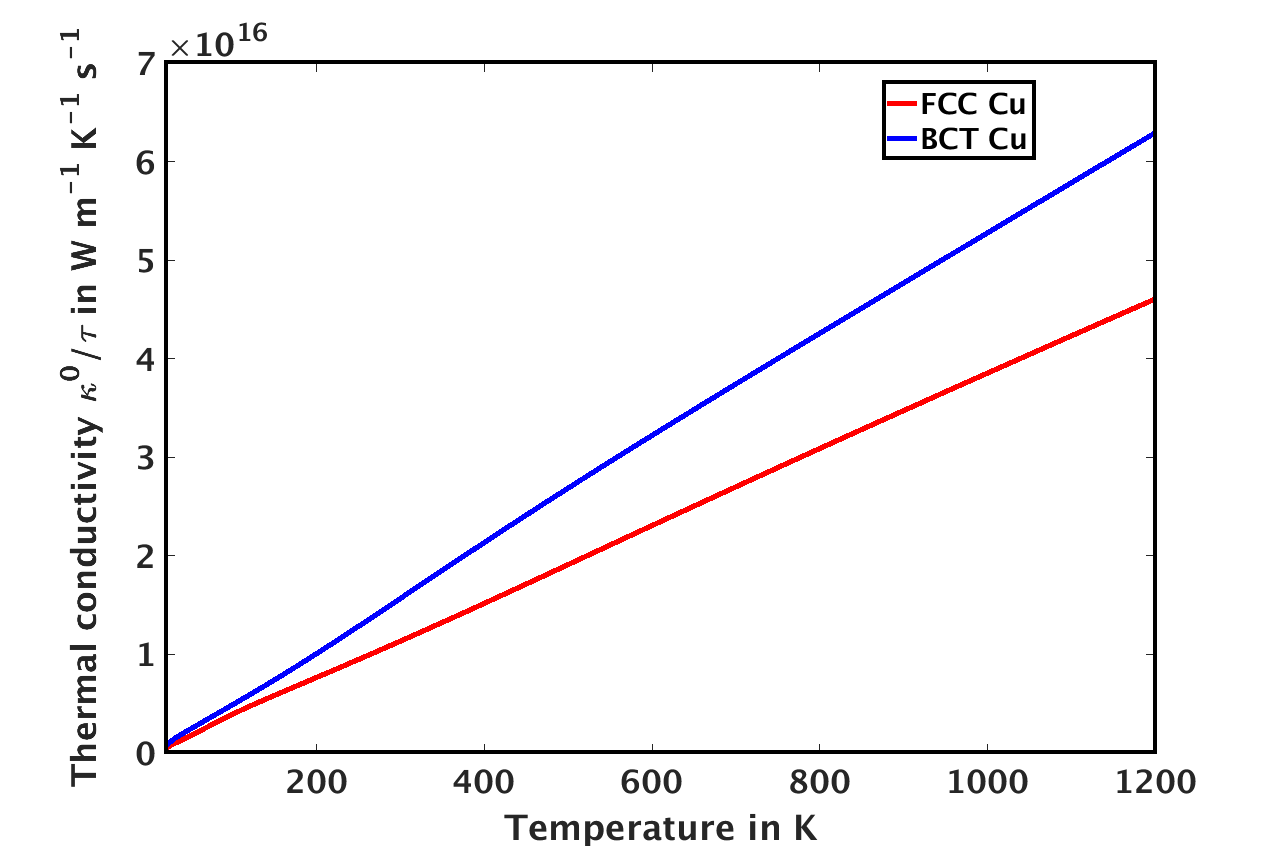}  
\caption{The electronic thermal conductivity of the two phases of Cu as a function of temperature. The high thermal conductivity of the BCT structure shows that it can be used as a good thermal conductor in electronic and spacecraft devices.}
\label{therm_conduc}
\end{figure} 

The calculated electronic band structures of FCC and BCT phases of Cu (Fig.~\ref{band_structure}) show that both the structures have finite density of states near the Fermi level and therefore they will show metallic behaviour. Based on the calculated band structures, we find out the electronic thermal conductivity of the two phases as a function of temperature for a fixed chemical potential using the semi-classical Boltzmann theory.~\cite{Boltzmann}  Though the thermal conductivity of a material consists of both electronic ($\kappa^0$) and phonon ($\kappa^l$) parts, BoltzTraP calculates only the electronic thermal conductivity $\kappa^0$ in terms of the relaxation time $\tau$, assuming that $\tau$ is constant and direction independent. From Fig.~\ref{therm_conduc}, it is seen that as temperature rises, the electronic thermal conductivity of both the structures increases linearly with increasing temperature.  
However, the slope is higher in case of the BCT phase. The high thermal conductivity of the BCT structure indicates that the found high pressure BCT structure can be used as a better thermal conductor compared to the known FCC Cu and may find its application in electronic devices, such as cooling chips or in spacecraft thermal control applications. 
\section{Conclusions}
The possibility of existence of a metastable body centered tetragonal phase for Cu has been established in this DFT study. These states may be achieved on simultaneous application of high temperature and pressure to a bulk sample of Cu which is possible through shock loading. The Helmholtz free energy distribution at a high pressure demonstrates more stability for the BCT phase compared to the FCC phase at high temperatures (which can be realized through shock loading). The BCT phase of Cu is observed to have high thermal conductivity compared to that of the FCC phase and may be beneficial for different high temperature engineering applications.

\acknowledgments
UD appreciates access to the computing facilities of the DST-FIST (phase-II) project installed in the Department of Physics, IIT Kharagpur, India. UD would like to acknowledge Dr. Monodeep Chakroborty for some useful discussions and the Ministry of Human Resource Development for research fellowship.

\end{document}